\documentclass[preprint,pre,a4paper,superscriptaddress,showpacs,nofootinbib]{revtex4}

\usepackage{amsmath}
\usepackage{amsfonts}
\usepackage{amssymb}
\usepackage{bm}
\usepackage{latexsym}

\usepackage{graphicx}

\usepackage{float}

\newcommand{\la}{\langle}
\newcommand{\ra}{\rangle}
\newcommand{\rvec}{\mbox{$\underline{r}$}}

\newcommand{\Fvec}{\mbox{$\underline{F}$}}
\newcommand{\kvec}{\mbox{$\underline{k}$}}
\newcommand{\Uvec}{\mbox{$\underline{U}$}}
\newcommand{\uvec}{\mbox{$\underline{u}$}}

\newcommand{\Lvec}{\mbox{$\underline{L}$}}
\newcommand{\Tvec}{\mbox{$\underline{T}$}}

\newcommand{\vvec}{\mbox{$\underline{v}$}}

\begin{document}
\sloppy

\title{Continuum limit of amorphous elastic bodies (III):\\
Three dimensional systems}

\author{F.~Leonforte}
\email[Email: ]{fleonfor@lpmcn.univ-lyon1.fr}
\affiliation{
Laboratoire de Physique de la Mati\`ere Condens\'ee et des Nanostructures,
Universit\'e Claude Bernard (Lyon I) \& CNRS, 43 Bvd. du 11 Nov. 1918,
69622 Villeurbanne Cedex, France}

\author{R.~Boissi\`ere}
\affiliation{
Laboratoire de Physique de la Mati\`ere Condens\'ee et des Nanostructures,
Universit\'e Claude Bernard (Lyon I) \& CNRS, 43 Bvd. du 11 Nov. 1918,
69622 Villeurbanne Cedex, France}

\author{A.~Tanguy}
\affiliation{
Laboratoire de Physique de la Mati\`ere Condens\'ee et des Nanostructures,
Universit\'e Claude Bernard (Lyon I) \& CNRS, 43 Bvd. du 11 Nov. 1918,
69622 Villeurbanne Cedex, France}

\author{J.P.~Wittmer}
\affiliation{Institut Charles Sadron, CNRS, 6, Rue Boussingault,
67083 Strasbourg, France}

\author{J.-L.~Barrat}
\affiliation{ Laboratoire de Physique de la Mati\`ere Condens\'ee
et des Nanostructures, Universit\'e Claude Bernard (Lyon I) \&
CNRS, 43 Bvd. du 11 Nov. 1918, 69622 Villeurbanne Cedex, France}

\begin{abstract}
Extending recent numerical studies on two dimensional amorphous bodies, we
characterize the approach of elastic continuum limit in three dimensional
(weakly polydisperse) Lennard-Jones systems. While performing a systematic
finite-size analysis (for two different quench protocols) we investigate the
non-affine displacement field under external strain, the linear response to
an external delta force and the low-frequency harmonic eigenmodes and their
density distribution.
Qualitatively similar behavior is found as in two dimensions. We
demonstrate that the classical elasticity description breaks down
below an intermediate  length scale $\xi$, which in our system is
 approximately $23$ molecular sizes. This length characterizes
the correlations of the non-affine displacement field, the
self-averaging of external noise with distance from the source and
gives the lower wave length bound for the applicability of the
classical eigenfrequency calculations.
We trace back the ``Boson-peak" of the density of eigenfrequencies
(obtained from the velocity auto-correlation function)
to the inhomogeneities on wave lengths smaller than $\xi$.
\end{abstract}


\maketitle

\section{Introduction.}
\label{sec:intro}

In a recent series of papers \cite{tanguy02,tanguy02b,tanguy04},
we investigated the elastic response of zero temperature
{\em two} dimensional (2D) amorphous systems.
Our studies were motivated by the idea that
such systems, although they appear perfectly homogeneous when
looking at the density field, may be described as heterogeneous
from the point of view of the theory of elasticity.
The basic reason for this failure is now well identified:
the underlying {\em hypothesis of affinity} of elastic deformations,
implicit in standard elastic theory, needs not to apply to a disordered system.
The relevant issue is therefore the scale above which a disordered,
glassy system can be considered as homogeneous from an elastic point of
view.

Obviously, this question is important for the vibrational spectrum
of such disordered systems; the excess of vibrational states at
intermediate frequencies in the spectrum (the so called ``boson
peak'') has previously been assigned to the existence of elastic
heterogeneities \cite{duval,novikov}. Moreover, the field of {\it plastic}
deformation of glassy materials, which has attracted considerable
attention recently \cite{maloneylemaitre,falklanger,cavaille,kabla,vandembroucq,picard} may be
expected to be related to elastic heterogeneities. Other points of
interest include the experimental evidence for dynamical
heterogeneities in deeply supercooled systems \cite{ediger},
which again could
be expected to give rise to ``frozen in'' heterogeneities in low
temperature systems.

Our previous studies were limited to 2D systems, as
this reduced dimension allows to carry out calculations on systems
with large  linear sizes using a limited number of particles.
These studies allowed us to establish,
for a standard computational model system,
the existence of a length scale that can reach a few tens
of particles, and below which classical elasticity breaks down.
Similar conclusions were reached  by Goldenberg and Goldhirsch
\cite{goldenberg2002}.  This breakdown is revealed by a number of
different diagnostics. Firstly, the so called Born expression for
elastic constants is found to give incorrect results at zero
temperature (where the fluctuation term does not contribute). This
failure can be traced back to the importance of a non-affine
contribution to the microscopic displacement field, while the
derivation of the Born formula assumes affine displacement at all
scales. The analysis of the correlation function of the non-affine
contribution to the displacement field reveals a length scale
$\xi$ over which this field is correlated, defining ``soft''
regions with large non-affine displacements. Second, the study of
low frequency vibrations in these model disordered systems shows
that the predictions of classical elastic theory are recovered
only for wave lengths larger than $\xi$, meaning the system is not
homogeneous from the point of view of elastic properties below
this scale.
More recently, it was shown that the response to a point force is
dominated by fluctuations for distances to the source smaller than
$\xi$ \cite{tanguy04}. Hence, $\xi$ characterises the self-averaging
of the noisy response within each configuration which let us to
call it the {\em self-averaging length}.
%
The pressure dependence of $\xi$ has also been investigated in 2D
demonstrating that the self-averaging length remains
``mesoscopic'' for low and moderate pressures,
typically of the order of $40$ particle sizes,
but decreases at large pressures \cite{pressure_results}.

An obvious question that arises is the extent to which these
results may depend, qualitatively or quantitatively, on the
dimensionality of space. Three dimensional (3D) systems, however,
are considerably more difficult to study than the 2D case. In
particular, if the elastic inhomogeneities take place on a length
scale comparable to what is observed in 2D, very large systems
have to be studied in order to reach the limit of elastically
homogeneous systems, i.e. $L\gg \xi$ where $L$ is the lateral size
of a cubic system and $\xi$ the size of inhomogeneities. Typical
numbers of the order of at least $10^5$ particles
should be considered, if one wants to use the same
tools and diagnostics in 2D and 3D 
systems.
Although a number of studies have appeared recently
\cite{depablo,nagel,rossi} pointing to the existence of elastic
inhomogeneities in various types of disordered systems,
all of them were realised for relatively small system sizes,
making a direct comparison with our previous results difficult.
In the same way, previous calculations of vibration modes in
3D systems have been limited to rather small
sizes \cite{schober,meshkov97,allen}.
To our knowledge, this work is the first one to explore systems with
lateral sizes that are appreciably larger than the expected scale
of elastic heterogeneities.

The aim of this work is therefore to characterise the elastic
behavior of large 3D systems, using the same computer model and
similar quench protocols as in our previous 2D studies.
They are summarised in section~\ref{sec:technicalities}.
In section \ref{sec:deformation}, we start by analysing the non-affine local
displacement field in cubic samples submitted to uniaxial elastic
deformation. From previous experience \cite{tanguy02,tanguy02b},
we know that this type of analysis is the most cost effective in revealing the
existence of inhomogeneities and their length scale. We then discuss
the elastic response to a point force (section \ref{sec:pointsource})
and corroborate why $\xi$ has been termed ``self-averaging length".
Vibrational properties at very low eigenfrequencies
-- obtained by diagonalization of the dynamical matrix --
are considered in section~\ref{sec:vibration} and
the density of eigenstates
-- computed by means of the finite temperature velocity
auto-correlation function --
in section~\ref{sec:dos}.
Our results are summarised in the last section.

\section{Description of systems and simulation procedures.}
\label{sec:technicalities}

The initial configurations and their preparation are very similar
to those described in Ref.~\cite{tanguy02} for the 2D case. 
In order to make the comparison 
as direct as possible, the same type of potential was
chosen, i.e. a slightly polydisperse Lennard-Jones potential
$U_{ij} (r)   = 4 \epsilon \left( \left( \sigma_{ij}/r\right)^{12} -
\left(\sigma_{ij}/r\right)^{6} \right)$
where the $\sigma_{ij}$ are taken uniformly distributed between
$0.8\sigma$ and $1.2\sigma$, corresponding to a polydispersity
index of $0.12$. This is expected to be enough to destabilise a
polydisperse crystal \cite{polydisp}, and indeed no sign of
crystallisation or demixing was observed in our simulations.
The interaction energy scale $\epsilon$ and the particle masses $m$
will be taken to be strictly monodisperse. In the following, 
we will adopt the units appropriate for this Lennard-Jones systems, 
i.e. the mean diameter $\sigma$ will be our unit of length 
(and generically described as the ``particle size''), 
while time will be expressed in units of $\tau= \sqrt{m\sigma^2/\epsilon}$.

We studied various systems at constant density, $\rho=N/L^3=
0.98$, which for our systems corresponds to a very low
hydrostatic pressure at zero temperature ($|P|\approx 0.2$). The lateral
size $L$ of the system was varied between $L=8$ and $L=64$
(corresponding to $N=500$ and $N=256 000$ particles). 

Disordered configurations are prepared by melting at high temperature 
($k_BT=2\epsilon$) an initially FCC configuration during $10^5$ 
Molecular Dynamics Steps (MDS) using constant temperature molecular
dynamics and velocity-Verlet integrator with a step size of $0.001\tau$. 
After the system was equilibrated, we
checked that no crystalline order remains, and we begin the production
run using two types of minimisation. The first one, called the ``fast'' quench,
uses a direct conjugate gradient minimisation until (according to
numerical tolerance) the zero temperature equilibrium state is reached. This
type of minimisation was implemented for all system sizes. The second
one, which we called the ``slow'' quench, is implemented for systems containing 
up to $N=62 500$ particles, corresponding to a lateral size $L=40$. 
The protocol used in this case consists in
equilibrating the liquid configuration at intermediate temperature 
($k_BT=1\epsilon$) and then cooling this by stages
($k_BT=5.10^{-1}\epsilon, 10^{-1}\epsilon, 5.10^{-2}\epsilon,
10^{-2}\epsilon, 5.10^{-3}\epsilon, 10^{-3}\epsilon$). 
At each stage, the system is ``aged'' (rather than equilibrated) 
during $10^5$ MDS. 
Finally, the zero temperature state is reached using conjugate
gradient minimisation from the last stage. Unless indicated otherwise, all the results discussed below refer to the fast quench procedure, which was carried out for all system sizes.

In order to obtain good quality statistics, this procedure was
repeated between $1$ and $10$ times, depending on the system size. 
The results presented in the following are 
averages over these different realisations of our amorphous systems.

\newpage
\clearpage
\section{Response to a macroscopic uniaxial deformation}
\label{sec:deformation}

\subsection{Computational procedure and non-affine
displacement fields}
\label{deform_nonaff}

In this section, we investigate the elastic behavior of zero
temperature cubic samples, prepared as described above, submitted
to an uniaxial traction. The procedure adopted is the following.
First, a global deformation of strain $\epsilon_{xx} \ll 1$ is
imposed on the sample by rescaling all the coordinates in an {\em
affine} manner.
Starting from this affinely deformed configuration, the system is
relaxed to the nearest energy minimum, keeping the shape of the
simulation box constant. As a result, a displacement of the particles
relative to the affinely deformed state is observed. This defines
the {\it non-affine} displacement field $\uvec(\rvec)$.
%


\begin{figure}[t]
\centerline{
\resizebox{7.5cm}{!}{\includegraphics*{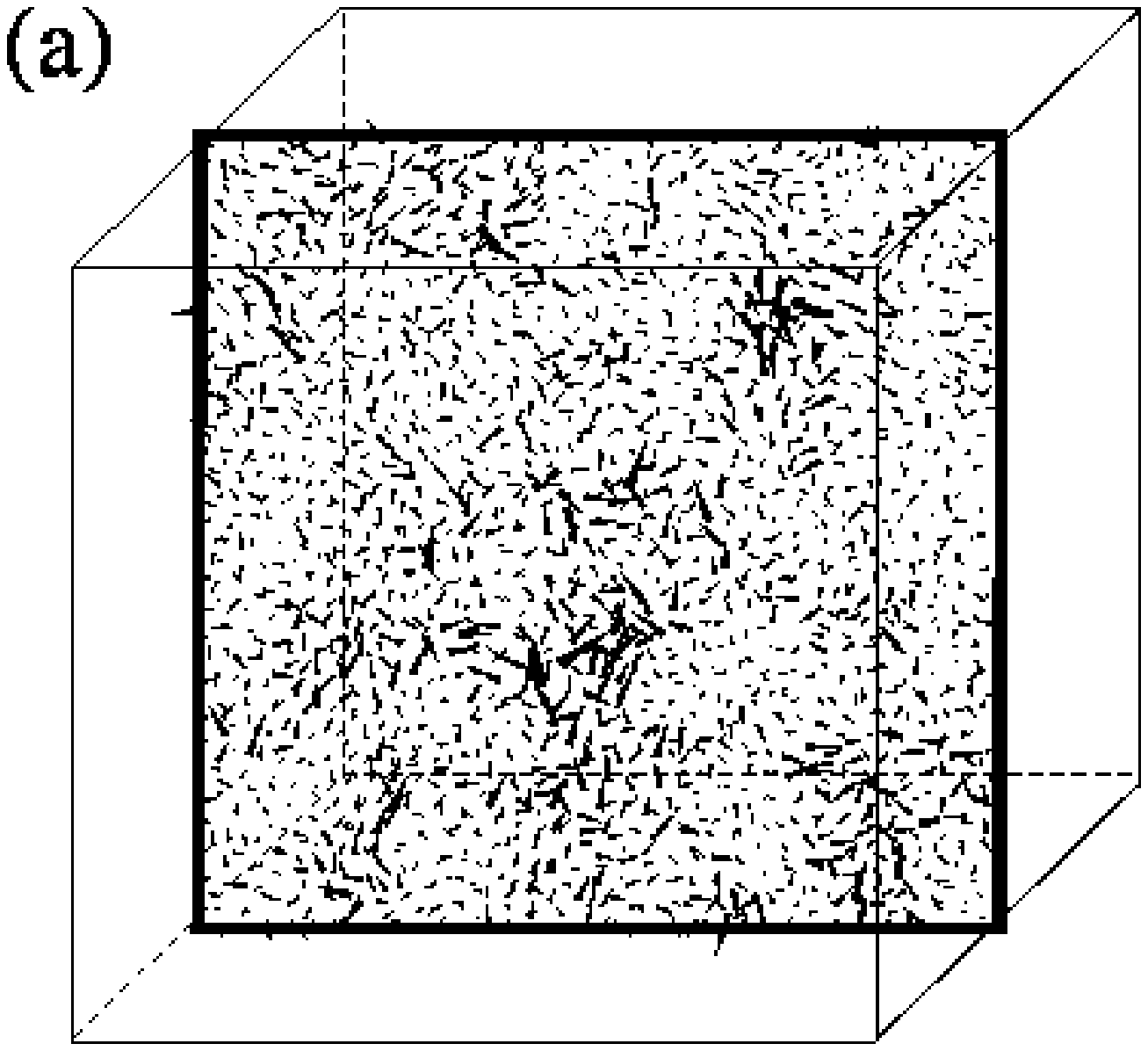}}
\hspace*{+0.5cm}
\resizebox{7.5cm}{!}{\includegraphics*{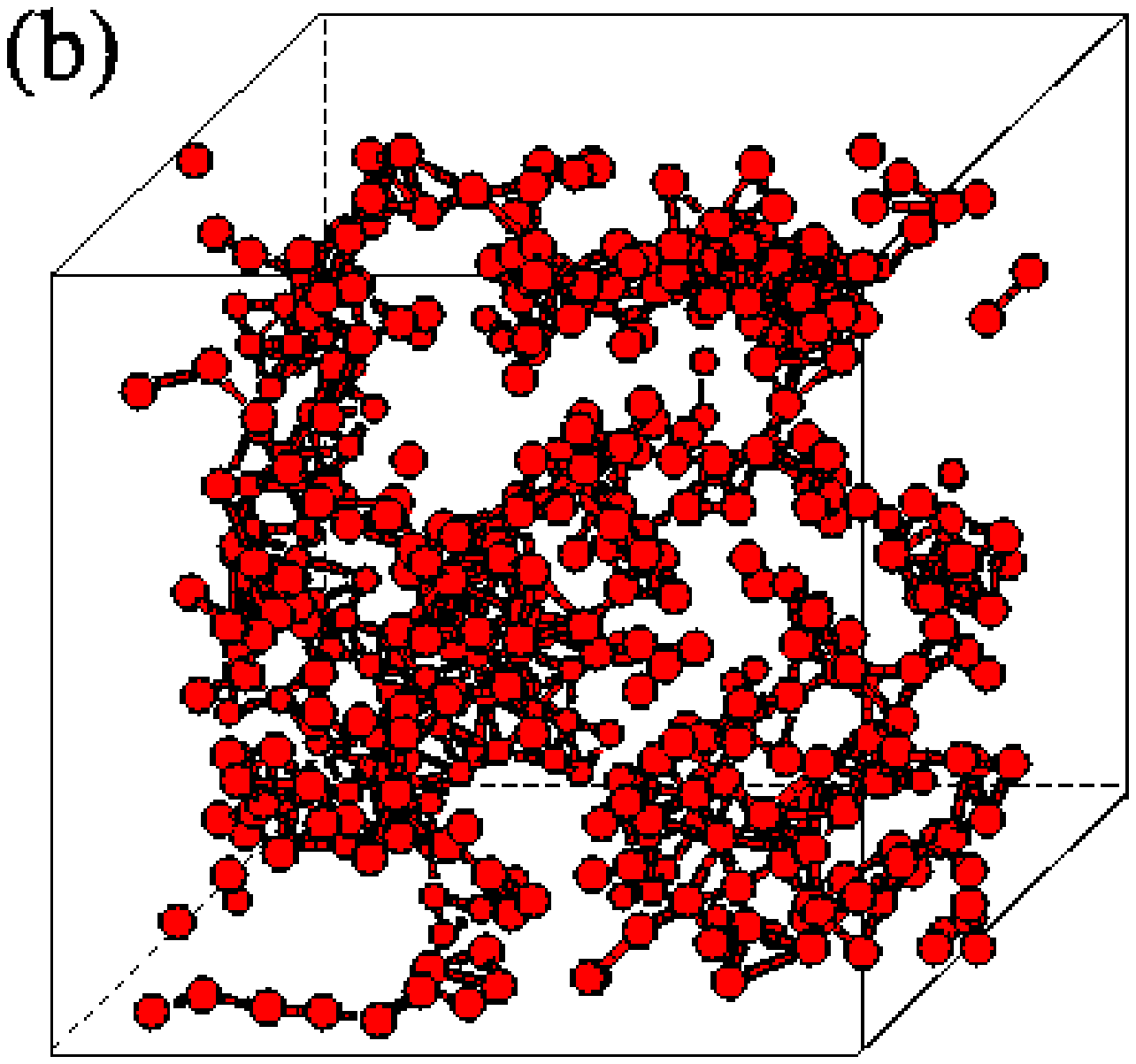}} }
\caption[]{
Non-affine part of the linear and reversible displacement field
$\uvec(\rvec)$ for the imposed macroscopic uniaxial strain
$\epsilon_{xx}=10^{-7}$
for a system containing $N=62500$ beads ($L=40$):
{\bf (a)}
projection on the ($x-z$)-plane for all particles close to the plane.
The length of the arrow is proportional to the displacement.
{\bf (b)}
all beads of the same configuration with the 10\% strongest
non-affine displacements.
(The short lines indicate beads with direct mutual interactions.)
This subset of beads is strongly spatially correlated on
short distances, however, it is homogeneously distributed and isotropic 
on larger scales.
\label{figsnap}}
\end{figure}

A typical example for this field in the {\em linear elastic}
response limit for a strain of $\epsilon_{xx} \approx 10^{-7}$)
is presented in the first panel of Fig.~\ref{figsnap}.
It displays a two dimensional projection of $\uvec(\rvec)$ in a 
plane containing the elongation direction for a system of size $L=40$.
(Note that projections on different planes are similar.)
Visual inspection of such snapshots suggests that non-affine fields
are {\em strongly} correlated over short and intermediate distances.
This impression is also confirmed by Fig.~\ref{figsnap}(b) where we
focus on the 10\% most mobile particles suggesting a connected cluster
of these strong displacements spanning the simulation cell.

\begin{figure}[t]
\centerline{
\resizebox{12.0cm}{!}{\includegraphics*{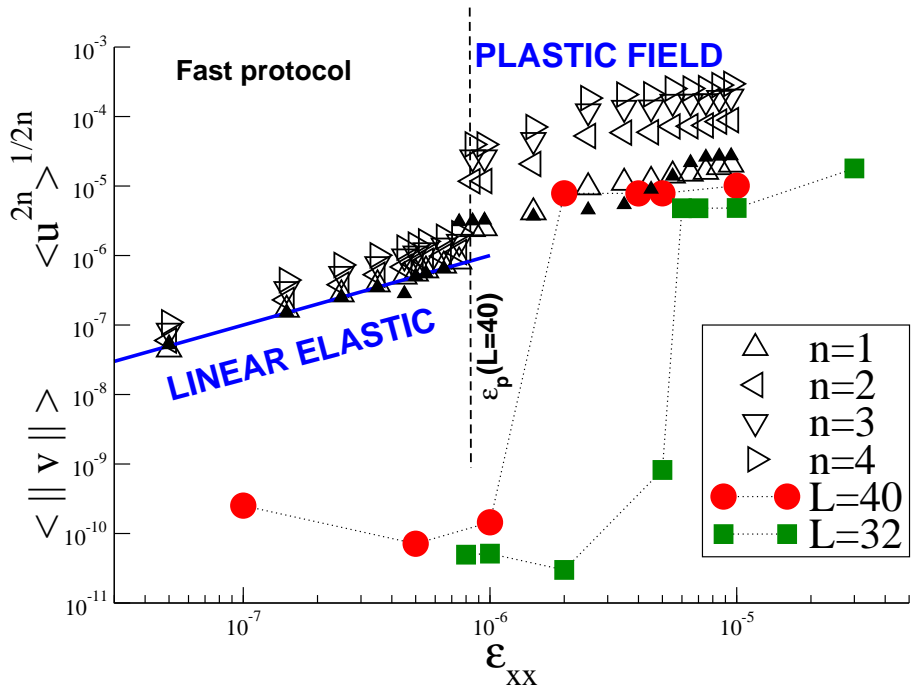}}}
\vspace*{-0.2cm} \caption[]{
Different moments of non-affine displacement field $\la \uvec^{2n}
\ra^{1/2n}$ for $n=1,2,3,4$ as a function of the imposed strain
$\epsilon_{xx}$ for systems of $L=40$ obtained by means of the
fast (open triangles) and the slow (full triangles) quench
protocol. Both protocols show very similar results. The bold line
on the left indicates the linear slope $\la \uvec^{2n} \ra^{1/2n}
\propto \epsilon_{xx}$. The vertical dashed line marks the limit
of elastic response $\epsilon_p(L) \approx 10^{-6}$ for $L=40$.
Also given is the residual plastic displacement field $\la || \vvec || \ra$
(obtained by reverse deformation back to the original macroscopic shape)
for $L=40$ and $L=32$ (full symbols). Residual fields below $10^{-9}$ are
due to numerical inaccuracies and the field can be considered
as reversible.
The sudden rise at $\epsilon_p$ for $L=40$ corresponds
nicely to the jump of the moments $\la \uvec^{2n} \ra^{1/2n}$.
Note that the plasticity threshold $\epsilon_p$ depends strongly on the system size.
\label{figu_strain}}
\end{figure}

In Fig.~\ref{figu_strain} we present the first moments
$\langle \left( \uvec(\epsilon_{xx}) \right)^{2n} \rangle^{1/2n}$
of the non-affine displacement field of system of size $L=40$
averaged over all particles of an ensemble. Both quench protocols
are very similar. (Only one moment is given for the slow protocol
for clarity.)
For $\epsilon \ll \epsilon_{p} \approx 10^{-6}$ (vertical line)
all moments are
(up to prefactors of order one) identical which demonstrates an unique
strain dependence for all beads. As one may also expect,
a linear strain dependence is found (bold line).
At $\epsilon_p$ the moments increase suddenly and differ over more
than an order of magnitude.
This suggests an inhomogeneous strain dependence of the non-affine
displacement field as will be discussed below
(Fig.~\ref{figpartratio}).
We note finally that the threshold $\epsilon_p$ decreases strongly
with the system size (not shown). Hence, linear response requires
much smaller deformations for large $L$.
%

\subsection{Plastic displacements and participation ratio}
\label{deform_plastic}

The elastic (reversible) character of the deformations was
checked by carrying out the reverse transformation and measuring
the residual displacement of the particles, $\vvec_i$, which
corresponds to plastic deformation. The moment $\la || v_i || \ra_i$
of the residual field is indicated in Fig.~\ref{figu_strain}.
For $\epsilon_{xx} \ll \epsilon_p$ it is negligible and the
deformation is, hence, elastic. Interestingly, elastic and
linear elastic regimes coincide essentially as can be seen
from the figure.
For larger strains the residual displacements increase
sharply over several orders of magnitude and coincide roughly
with the $n=1$ moment of the non-affine displacements. This shows
that, for $\epsilon_{xx} > \epsilon_p$, the non-affine displacements are mainly due to plastic rearrangements.

\begin{figure}[t]
\centerline{\resizebox{12cm}{!}{\includegraphics*{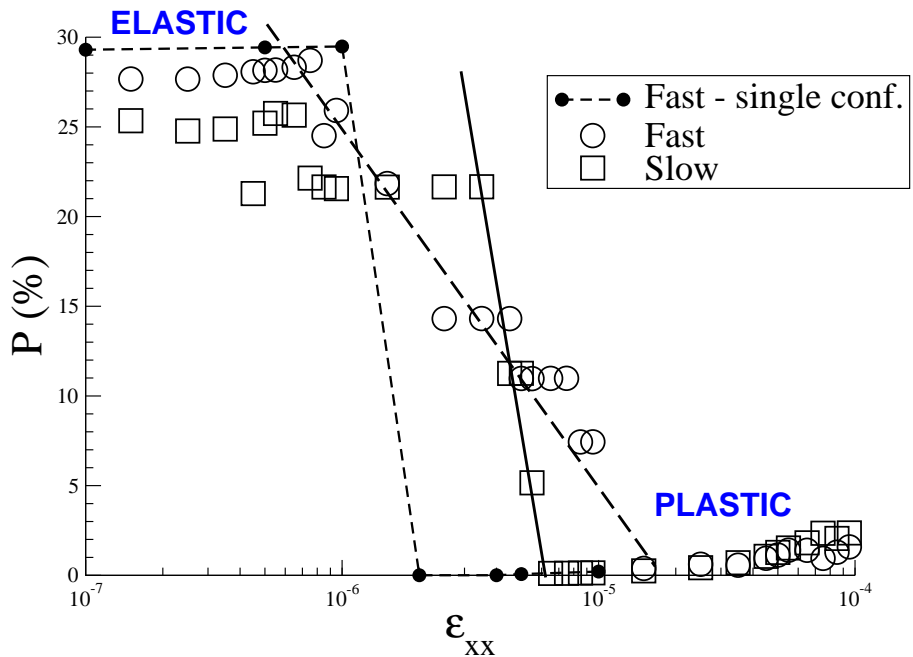}}}
\vspace*{-0.2cm} \caption[]{
Participation ratio of the non-affine displacement field in a 3D system
containing $62500$ particles ($L=40$), as a function of the uniaxial
strain $\epsilon_{xx}$.
Results for both fast and slow quench protocols averaged
over eight and five configurations respectively.
The given lines are guides to the eye.
In the case of the fast protocol, a single configuration is also shown, for comparison with the averaged case.
\label{figpartratio}}
\end{figure}

In view of the potential relationship with plastic deformation, it
is interesting to investigate in some detail the spatial features
of the non-affine displacement field. Qualitatively, this can be
achieved by representing, as shown in Fig.~\ref{figsnap}(b), the
particles that have the 10\% largest non-affine displacements.
This picture shows that the non-affine field is rather
delocalized, with the cluster formed by the most mobile particles
spanning the entire simulation cell. A more quantitative view can
be obtained by calculating the participation ratio for the
non-affine displacement, defined by
\begin{equation}
\label{eq:pr}
P=\frac{1}{N}  \frac{(\sum_i \uvec_i^2)^2}{\sum_i (\uvec_i^2)^2}
\end{equation}
This participation ratio is shown in Fig.~\ref{figpartratio}
for both quench protocols and $L=40$.
(A similar participation ratio may be calculated for the {\it residual}
plastic field. However in this last case, the ratio at small $\epsilon$ is due
to numerics and at high strain it is identical to the participation
ratio of the non-affine field.)
%
%
Obviously, for sufficiently small deformations the displacements
must depend identically for {\em all} beads on the applied strain
and $P$ has to become constant.
As anticipated in Fig.~\ref{figu_strain}, the presented data shows
that this coincides with the linear elastic regime where all moments of
the non-affine field are similar and the residual plastic field
can be neglected.
\footnote{ Note that the jumps in the plateau value at small strains (below $10^{-6}$) in Fig.~\ref{figpartratio}, specifically for the slow quench protocol, may be attributed to numerical inaccuracies due
to the small displacement fields which have been compared.}
The central point is here that the plateau value of the participation ratio is large (about 25\%) indicating that the elastic non-affine displacement involves a substantial fraction of the particles.
%
%
When the deformation exceeds the plastic threshold $\epsilon_p$,
however, the participation ratio falls rapidly, indicating that a
plastic deformation proceeds via well {\em localized} events. 
\footnote{It should be stressed that the more gradual decay of the participation ratio, in the fast protocol case (Fig.\ref{figpartratio}), is due to ensemble average. It describes essentially the probability that no jump has occurred for smaller $\epsilon$ values, and it is thus related to the distribution of the plastic thresholds $\epsilon_p$ from one configuration to the other. In the case of a single configuration, the decay of the participation ratio is more sudden, and can be compared with the averaged decay in the slow protocol case, where the distribution of $\epsilon_p$ is narrower. This is, in fact, the main difference observed between the two quench protocols: the plastic threshold $\epsilon_p$ is well defined in the slow quench protocol case, but it is largely distributed (over more that one decade) in the fast quench protocol case. Note however that we do not have, at this stage, a sufficient number of configurations to compute more precisely this distribution. } 
This is the first main result of this work.
The implication from this difference in behavior is that the
localized events occurring in plastic deformation cannot be
directly inferred from the general {\em pattern} of non-affine
displacements. This does not mean that plastic displacements and
strong non-affine elastic displacements are completely
uncorrelated.

In other words, energy barriers (which are relevant for plastic
deformation) are not directly related to the local curvature of
the energy minima \cite{lacks}.
Interestingly, the main influence of performing a slow quench seems to
be that the plasticity limit is increased, meaning that the system has
been brought to a slightly more stable configuration with higher energy
barriers without, apparently, changing measurably the local curvature
of the energy minima. In fact, properties such as the vibrational modes
discussed below, are much less affected by the quench protocol.

In the reminder of this paper, we focus on the {\em linear}
elastic response. We normalise the non-affine field by
its second moment, i.e., $\uvec_i$ is replaced by
$\uvec_i/\la \uvec^2 \ra^{1/2}$, in order to consider
a strain independent reduced field.

\subsection{Hydrodynamic limit: Lam\'e coefficients}
\label{deform_lame}

\begin{figure}[t]
\centerline{\resizebox{12cm}{!}{\includegraphics*{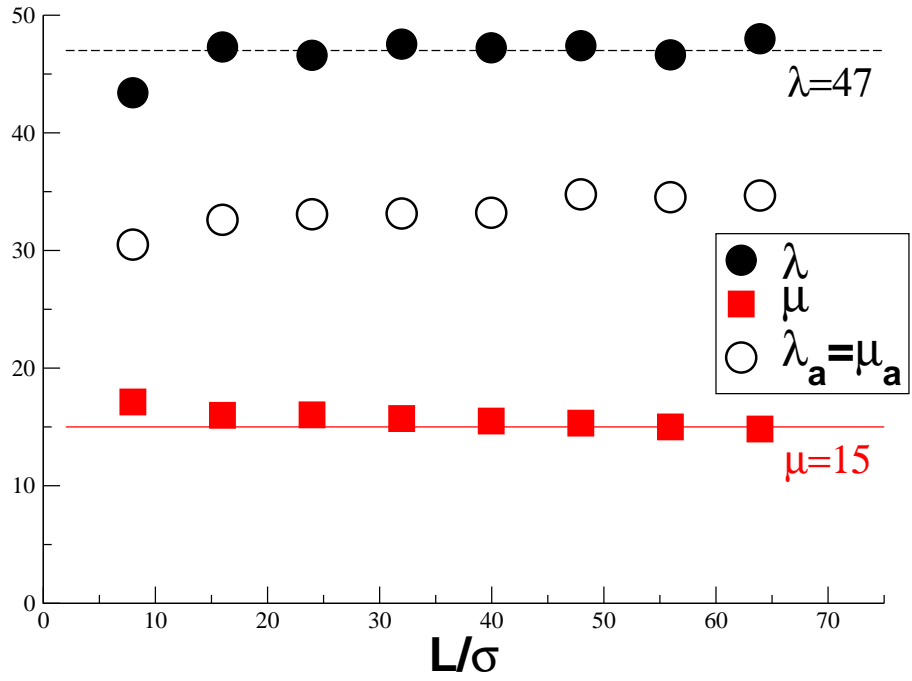}}}
\vspace*{-0.2cm} \caption[]{
Lam\'e coefficients $\lambda$ (spheres) and $\mu$ (squares) {\em vs.}
system size $L$.
Full symbols correspond to the direct measurement using Hooke's
law, open symbols are obtained from Eq.~(\ref{eq:lame})
which supposes {\em affine} deformations (Born term).
The effect of system size is weak.
For large boxes we get $\mu \approx  15$ and
$\lambda \approx 47$.
The coefficients relying on a negligible non-affine field differ
by a factor as large as 2 from the true ones. Clearly, a
calculation taking into account the non-affine character of the
displacement is necessary for disordered systems. \label{figlame}}
\end{figure}

We turn now to the calculation of the Lam\'e coefficients
$\lambda$ and $\mu$, which characterize the elastic behavior of an
isotropic medium in 3D \cite{landau}. Our results for these
coefficients as a function of system size are shown in
Fig.~\ref{figlame}. This figure compares two different ways of
obtaining the  coefficients. $\lambda_a$ and $\mu_a$ are obtained
under the assumption that the non-affine contribution to the total
displacement field of the particles is negligible. They are simply
the Born estimates, that can be, in a system with pairwise
interactions, computed from the reference configuration by
carrying out a simple summation over all pairs of interacting
particles (see for example ref.~\cite{maloneylemaitre})
\begin{equation}
\lambda_a=\mu_a= \frac{1}{L^3}\sum_{i,j} \left(U^{\prime\prime}(r_{ij}) -
\frac{1}{r_{ij}} U^{\prime}(r_{ij})\right) \frac{x_{ij}^2
y_{ij}^2}{r_{ij}^2}
\label{eq:lame}
\end{equation}
where $U$ is the interaction potential. The second estimate
corresponds to the ``true'' value of the elastic coefficients,
obtained by computing the actual stress in the sample {\it after}
the relaxation that introduces the non-affine part of the
displacement field. The averaged stress is computed using the usual
microscopic expression
\begin{equation}\label{stress}
    \overline\sigma_{\alpha\beta} = \frac{1}{L^3} \sum_{i,j}(\rvec_{ij})_\alpha
    (\Fvec_{ij})_\beta
\end{equation}
where the Greek indices refer to cartesian coordinates, and
$\Fvec_{ij}$ is the force between particles $i$ and $j$.

The Lam\'e coefficients are then obtained from the standard
formulae $\overline\sigma_{xx} = (\lambda+ 2\mu) \epsilon_{xx}$ and
$\overline\sigma_{yy}= \lambda \epsilon_{xx}$ for a deformation tensor
which has only an $\epsilon_{xx}$ component
($\epsilon_{xx}$ is here the global deformation imposed on the sample).
For larger systems, we obtain $\mu \approx  15$ and $\lambda \approx 47$.
Hence, we find that the true values of $\lambda$ and $\mu$ differ
considerably from the Born estimates which indicates the importance
of non-affine displacements in determining the stresses in the material.
This contribution tends to lower the shear modulus $\mu$, and to increase
the coefficient $\lambda$.
From the measured values of $\lambda$ and $\mu$ we get
a bulk compression modulus $K=\lambda+2\mu/d \approx 57$,
a Young modulus $E=8K\mu/(3K+\mu) \approx 37$ and
a Poisson ratio $\nu = (3K-2\mu)/2(3K+\mu) \approx 0.4$
\cite{landau}.
Remarkably, the bulk modulus $K$ would be correctly
predicted by the Born calculation. This means that the non-affine
part of the deformation does not contribute significantly to the
increment in the {\em isotropic} pressure under compression
or traction, but is mainly associated with shear deformations.
(This point will be further elucidated below when we will
discuss Fig.~\ref{figSTLk}.)
Such a situation would be natural in high pressure systems,
in which the repulsive inverse power part of the potential
dominates the interaction, and compression can be accommodated
by an affine rescaling of all coordinates.
It is, however, less expected in our low pressure systems.

\subsection{Correlations in the non-affine displacement field}
\label{deform_corr}

The preceding results call for a more thorough analysis of the
correlations of the non-affine displacement field which apparently
can not be neglected for macroscopic quantities and should therefore
be even more relevant for finite wave length properties.
%

\begin{figure}[t]
\centerline{
\resizebox{12.0cm}{!}{\includegraphics*{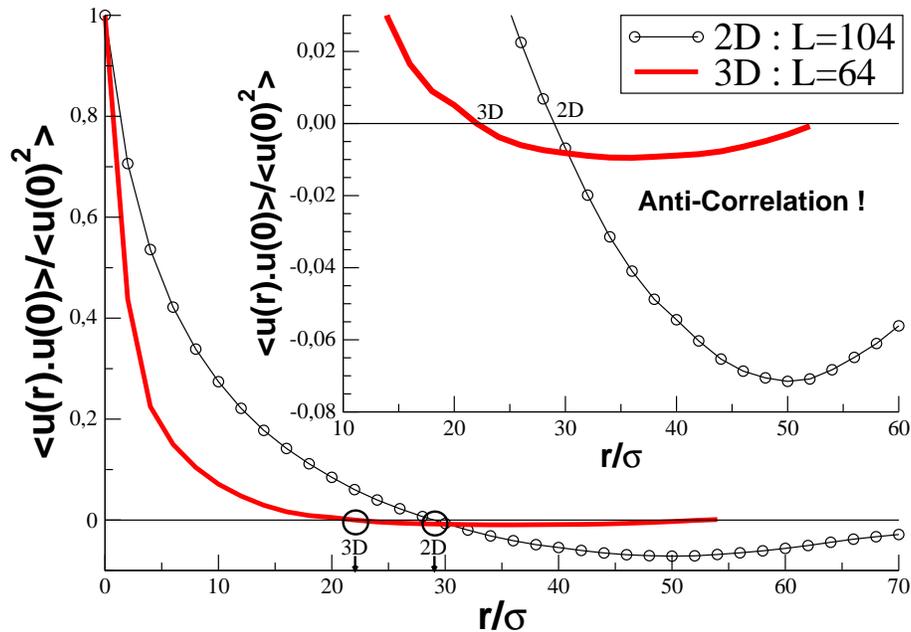}}}
\vspace*{-0.2cm} \caption[]{
Correlation function $C(r)$ of the non-affine deformation field as
a function of the distance between pairs of beads $r$. The
correlation function is averaged over all possible pairs. The full
line corresponds to a 3D sample containing $256 000$ particles
($L=64$). Note that the "fast" quench protocol and the "slow" quench 
protocol (not shown here) give the same curve. 
Data from \cite{tanguy02} for a 2D system of linear size
$L=104$ is shown for comparison (line with open dots). In the 2D
case, results are averaged over 30 configurations. The data are
plotted in linear scale. Inset: Enlargement of the
anti-correlation regime. Note the {\em negative} -- although weak
-- correlation of the 3D correlation function for $r>23\sigma$.
\label{figCr}}
\end{figure}

Following Ref.~\cite{tanguy02,tanguy02b}, the non-affine correlation field
can be analyzed by computing the correlation function $C(r) \equiv
\langle {\uvec(\rvec) \cdot \uvec(0)} \rangle$. (The averages are
taken over all pairs of monomers $(i,j)$ being a distance $r$
apart.) As can be seen in Fig.~\ref{figCr}, a decay over a typical
length of 23 particle sizes is observed (bold line), before the
correlation function exhibits  a {\em negative} tail (see inset).
The 2D case (disks) is also included for comparison. Qualitatively
similar behavior is found. The anti-correlation can be associated
visually in 2D with the solenoidal character of the non-affine
field \cite{tanguy02,tanguy02b}. The organisation of the non-affine
deformation in ``vortices'' is less obvious in 3D (see inset) as
manifested by the about seven times weaker amplitude of the
negative tail. However, this description can be compared with the direct visualization of Fig.~\ref{figsnap}, where the intricate structure of the vortices is shown. See also Fig.~\ref{figSTLk} below.

\begin{figure}[t]
\centerline{
\resizebox{13.0cm}{!}{\includegraphics*{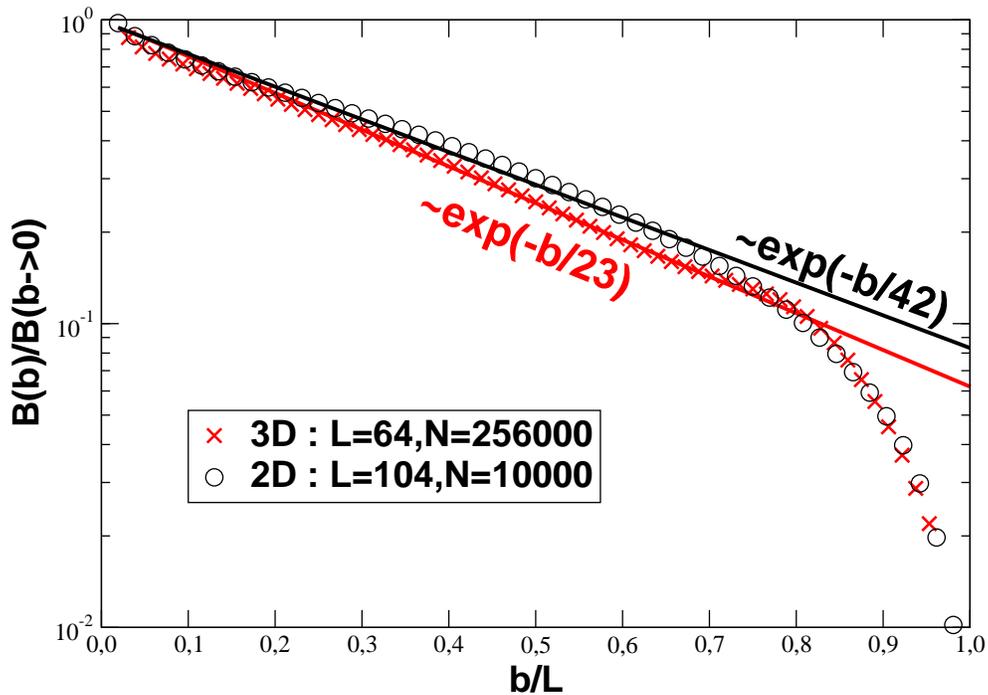}}}
\vspace*{-0.2cm} \caption[]{
The (normalized) magnitude $B(b)$ of the non-affine field averaged
over a volume element of lateral size $b$ is traced as a function
of $b/L$. The plot confirms that the correlations decay with a
characteristic length $\xi=23$ for 3D and $\xi=42$ for 2D
(spheres, for 10 configurations) respectively. \label{figBb}}
\end{figure}

That the displacement field is indeed correlated over a size
$\xi \gg \sigma$ is further elucidated in Fig.~\ref{figBb}.
Here we consider the systematic coarse-graining of the
displacement field
\begin{equation}
\underline{U}_j(b) \equiv \frac{1}{N_j} \sum_{i \in V_j} \uvec(\rvec_i)
\label{eq:Ucoarse}
\end{equation}
of all $N_j$ beads contained within the cubic volume element $V_j$
of linear size $b$. In the figure, we have plotted the
(normalized) correlation function $B(b) \equiv (\langle
\underline{U}_j(b)^2 \rangle_j/\langle \uvec^2 \rangle)^{1/2}$
versus the size of the coarse-graining volume element $b$
(normalized by $L$). Two and three dimensional systems are again
compared. In both cases we find an {\em exponential} decay which
is well fitted by the characteristic scales $\xi\approx 23$ for 3D
and $\xi\approx 42$ for 2D. Apparently, $\xi$ is similar to the
distance where $C(r)$ becomes anti-correlated. Note that the total
non-affine displacement field of the box must vanish -- since the
 center of mass of the system is fixed-- and therefore $B(b) \rightarrow
0$ for $b/L \rightarrow 1$. This sum rule explains the curvature
in the
data and the sharp 
cut-off on the right hand side of the figure.
The nice agreement of both estimations of $\xi$ demonstrated in
the Figs.~\ref{figCr} and \ref{figBb} and the similar behavior in
both dimensions is the second central result of this work.


\begin{figure}[t]
\centerline{\resizebox{13cm}{!}{\includegraphics*{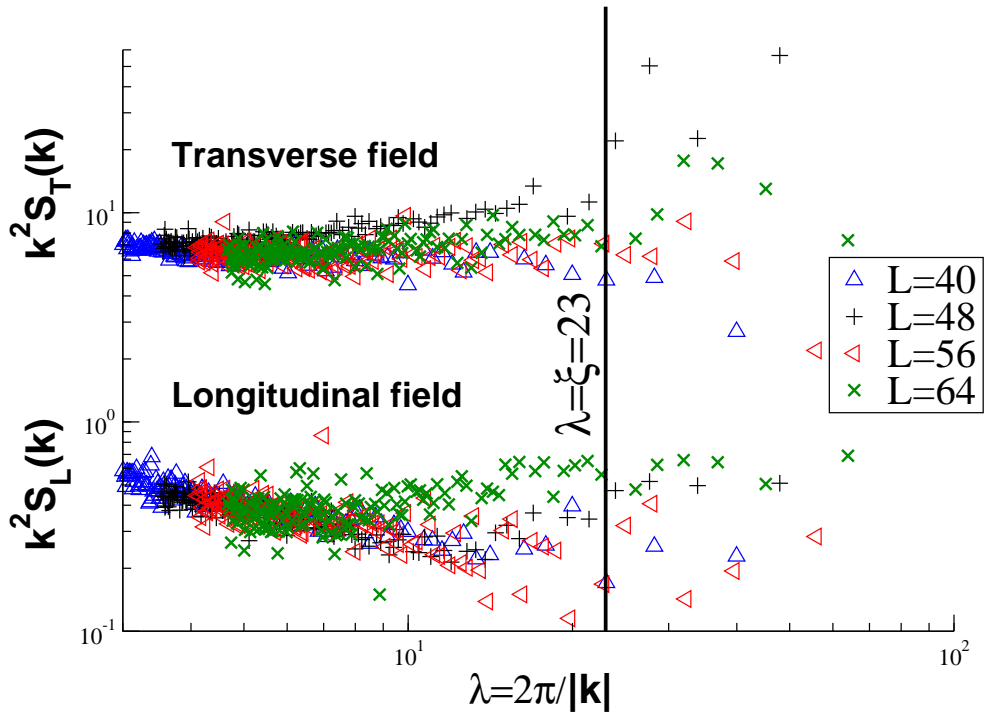}}}
\vspace*{-0.2cm} \caption[]{
The squared amplitudes
of the Fourier transforms $k^2 S_L$ and $k^2 S_T$
for the div (lower data) and curl (upper data) of the non-affine
deformation field (see Eq.~(\ref{eq:STk}))
plotted as functions of the wave length $\lambda = 2\pi/||\kvec||$.
Different system sizes, as indicated in the figure, have been
included to demonstrate that $S_L$ and $S_T$
are essentially system size independent for small $\lambda$.
Note that the statistics deteriorates for large $\lambda$.
\label{figSTLk}}
\end{figure}

More systematically, the displacement field can be investigated in
Fourier space
$\Uvec(\kvec) \equiv \sum_{i=1}^N \uvec(\rvec_i) \exp(i \kvec \cdot \rvec_i)$.
Apart the normalisation factor $1/N_j$, this is close to the
volume element coarse-graining method, Eq.~(\ref{eq:Ucoarse}). As
usual, the fluctuations can be directly computed from $S(k) \equiv
\frac{1}{N} \langle || \Uvec(\kvec) ||^2 \rangle$ where the
average is taken over all wave vectors with $k = ||\kvec||$. Our
results for $S(k)$ are not presented since we prefer to discuss
below in Fig.~\ref{figSTLk} the longitudinal and transverse
contributions, $S_L(k)$ and $S_T(k)$. Note also, that $S(k)$ is
related to the real space coarse-graining correlation function
$B(b)$ of linear size $b=2\pi/k$ given in
Fig.~\ref{figBb}.\footnote{ The wave vector $\kvec$ must be chosen
commensurate with the periodic simulation box. For this reason
fewer values of the wave length $\lambda = 2\pi/||\kvec|| =
L/\sqrt(n^2+m^2+l^2)$ ($n$, $m$ and $l$ being integers) can be
computed. This is the principle disadvantage of the Fourier
transformation of the non-affine field for computational studies
in 3D -- where the system size is necessarily rather limited --
compared to the real space coarse-graining method which allows
also a continuous variation of the box length $b$.
Moreover, it is easy to show that $B^2(b)$ is the average over a volume $b^d$ of the correlation function $C(r)$ whose Fourier transform is related to $S(k)$. Thus a clearly shown size dependence of $B(b)$ can be masked in the power law decay of $S(k)$ (at large $k$). 
}
Interestingly, it can be readily shown that
\begin{equation}
S(k) = \frac{1}{N} \sum_{i,j}
\exp(i \kvec\cdot (\rvec_{i}-\rvec_{j})
\langle \uvec(\rvec_i) \uvec(\rvec_j) \rangle
= \langle u^2 \rangle +
\int  d\rvec \exp(i \kvec \cdot \rvec) C(r) \rho g(r)
\end{equation}
with $C(r)$ being the displacement correlation function
discussed in Fig.~\ref{figCr} and $g(r)$ the standard
pair distribution function. Since $g(r)$ becomes rapidly
constant it follows that $S(k)$ corresponds essentially
to the Fourier transform of $C(r)$.
This demonstrates that (very roughly) $C(r)$ and $B(b)$
(shown in Fig.~\ref{figCr} and \ref{figBb} respectively)
are related by Fourier transformations and explains why similar
characteristic sizes $\xi \gg \sigma$ have been obtained.

We demonstrate now that the non-affine displacement field in 3D
is indeed of predominantly {\em solenoidal} nature.
This has been anticipated by our previous studies on 2D systems
\cite{tanguy02} and by the values of the elastic moduli $\mu$
and $K$ discussed in Sec.~\ref{deform_lame}.
The transverse and longitudinal contributions to the displacement
field can be numerically obtained by computing
\begin{equation}
\Tvec(\kvec) \equiv -\frac{1}{k^2} \kvec \wedge
\left(\kvec \wedge \Uvec(\kvec) \right)
\mbox{  , }
\Lvec(\kvec) \equiv \frac{1}{k^2} \kvec
\left( \kvec \cdot \Uvec(\kvec) \right)
\end{equation}
Obviously, $\Uvec=\Tvec+\Lvec$,
$\kvec \cdot \Uvec = \kvec \cdot \Lvec$,
$\kvec \wedge \Uvec = \kvec \wedge \Tvec$
and $\kvec \cdot \Tvec = \kvec \wedge \Lvec =0$.
The norms of these quantities, for instance
\begin{equation}
\label{eq:STk}
k^2 S_T(k) \equiv \frac{k^2}{N} \langle ||\Tvec(\kvec)||^2 \rangle
= \frac{1}{N} < ||\sum_i \kvec \wedge \uvec(\rvec_i) \exp(i\kvec
\cdot \rvec_i)||^2 >
\end{equation}
and similarly for the longitudinal part $k^2 S_L(k)$, are the
Fourier transforms of $\underline{\nabla} \wedge  \uvec$ and
$\underline{\nabla} \uvec$. They are plotted in Fig.~\ref{figSTLk}
as functions of $\lambda = 2\pi/k$. We find that the longitudinal
contribution (data points at the bottom) is about 10 times smaller
than the transverse one. Note that all data points of different
system sizes collapse well for wavelengths corresponding to the
non-affine displacement regime, $\lambda < \xi$.
We find that $k^2 S_T(k)$ is more or less constant while $k^2
S_L(k)$ decreases weakly. Since $\Tvec$ and $\Lvec$ are orthogonal
this yields $S(k) = S_T(k) + S_L(k) \approx S_T(k)$ and, hence,
the algebraic relation $S(k) \propto 1/k^2$ for large $k$.
\footnote{see previous footnote.}
Unfortunately, for larger $\lambda$ the statistics deteriorates due to the
smaller number of wave vectors which can be considered. Our data may suggest
that $S_T(k)$ increases for $\lambda > \xi$, however, new data with larger
boxes and with better statistics is warranted to confirm this.

\newpage
\clearpage
\section{Self-averaging of the response to a point force }
\label{sec:pointsource}

\begin{figure}[t]
\centerline{\resizebox{13cm}{!}{\includegraphics*{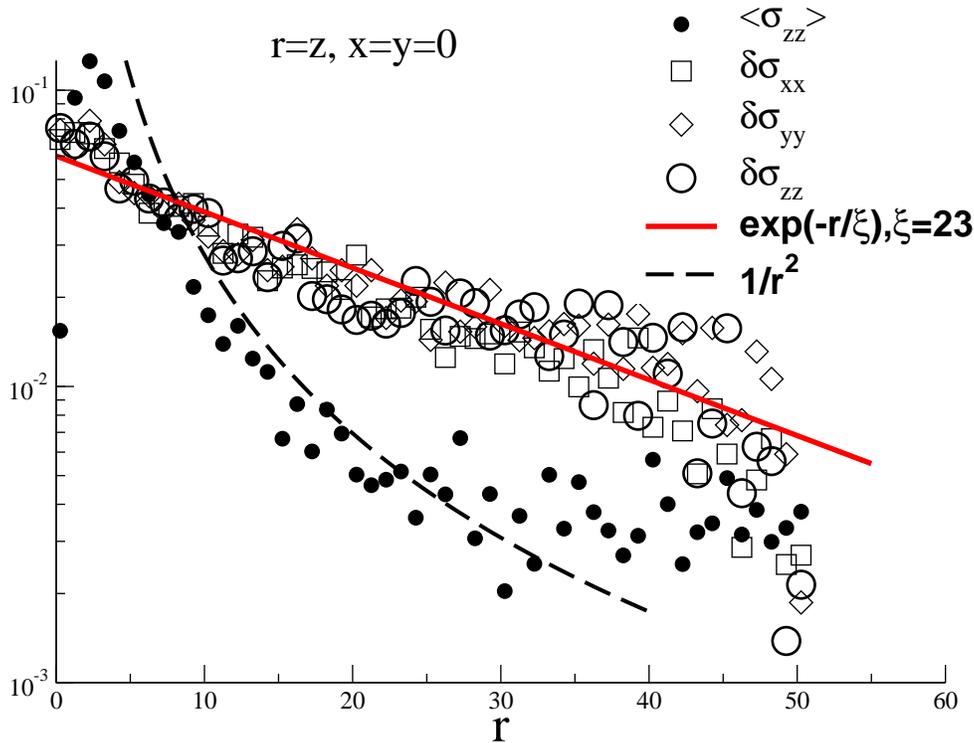}}}
\vspace*{-0.2cm} \caption[]{
Comparison of the incremental stress fluctuations
$\delta\sigma_{\alpha\beta}\equiv(<\sigma^2_{\alpha\beta}>-<\sigma_{\alpha\beta}>^2)^{1/2}$
with the mean vertical component of normal stress $<\sigma_{zz}>$ for
the ensemble average discussed in text, measured along the vertical line
($x=y=0$). The mean stress decreases essentially as $1/r^2$, as expected
(without logarithmic corrections) in 3D for positions far from
the source and the fixed walls. (Note that $<\sigma_{zz}>$
has to decrease less rapidly close to the walls, $r\approx L/2$.)
In contrast, the fluctuations decay exponentially over the whole
available system.
The characteristic length scale is similar to the size 
$\xi \approx 23 \sigma$ of the correlated non-affine displacements.
\label{figselfaverage}}
\end{figure}

In 2D,
we showed previously \cite{tanguy04} that the
deviations from continuum elasticity at small scales could be
revealed by studying the response of the system to  a localised
force. The same effect is illustrated for 3D in 
Fig.~\ref{figselfaverage}. This plot is obtained as follows. A small
force is applied to the particles contained in a small region of
space (sphere of diameter $4$). The force is applied in the $z$
direction, and its magnitude is chosen small enough to remain in
the (linear) elastic region ($F_z=10$). In order to maintain global force
balance, the system has periodic boundary conditions in the $x$
and $y$ directions, but is immobilised by two fixed walls in the
$z$ direction. 
The dimensions of the simulation cell, which
contains $N=165000$ particles, are $L_z\simeq 105$ in the $z$
direction, $L_x=L_y\simeq40$
in lateral directions.
The stress tensor is then measured within
the sample using the standard Kirkwood definition
\cite{goldenberg2002} calculated in small rectangular boxes of fixed 
size $(3,3,5)$ centered in $(x,y,z)$. 
Those boxes are displaced in all 
the material by unit steps of one in the three directions. For each
step the six components of the stress tensor are calculated and 
averaged on a statistical ensemble of $200$ configurations. Such an
ensemble is obtained by taking $10$ independent configurations, and
for each configuration by changing the position of the point source
within the sample reindexing the configuration in such a way that
the origin of the source still is placed in the midplane,
equally distant from the two fixed walls. 
%
In the absence of the external force, the local stresses in amorphous 
systems are usually nonzero (``quenched stresses''). The relevant 
quantity that defines the response to an external force are therefore 
the incremental, rather than total, local stress tensor. 
Once such a calculation done, the quantity of interest is the 
fluctuations of the stress tensor in the statistical ensemble 
\cite{tanguy04}.

In Fig.~\ref{figselfaverage}, both the average value and the 
fluctuations of the stress tensor are shown. 
The average response is compared to the prediction of continuum
elasticity, which appears to be perfectly obeyed {\it on average}
even very close to the source. This average response exhibits the
$1/r^2$ decay characteristic of the Green function of classical
elasticity in 3D. 
However, up to length scales of  50 the fluctuations are considerably 
larger than the average value of the stress. 
The fluctuations, on the other hand decay {\em exponentially} away 
from the source, with a characteristic length $23$ -- the same 
value as obtained from the correlation functions discussed in the 
previous section.
In fact, the {\em relative} stress fluctuations,
for instance $\delta\sigma_{zz}/\langle \sigma_{zz} \rangle$, 
scale like $\exp(-r/\xi) r^2$ and show non-monotoneous behaviour
(not shown). 
They {\em increase} first up to $2\xi$ (due to the decreasing mean stress), 
but decrease ultimately exponentially. With other words, {\em self-averaging} 
(within every configuration) of the noisy signal occurs on distances set by 
$\xi$ which we call, hence, the {\em self-averaging length}.
\footnote{Absolutely similar behaviour has also been found in 2D systems 
\cite{tanguy04}. There the relative fluctuations scale like 
$\exp(-r/\xi) r$, i.e. exponential screening sets in for distances 
$r \gg \xi$. Since $\xi$ is about twice larger in 2D it turns out 
that about the {\em same} linear distance is required in both dimensions 
for self-averaging to become effective. This suggests that
if different dimensions $d$ are compared one should rather use 
the notion ``self-averaging length" for $(d-1) \xi$.
}
Unfortunately, our simulation boxes are yet too small to illustrate
more clearly the exponential decay of the relative noise for 
$r\gg 2\xi \approx 50$.

\newpage
\clearpage
\section{Low frequency equilibrium vibration modes}
\label{sec:vibration}

\begin{figure}[t]
\centerline{\resizebox{13cm}{!}{\includegraphics*{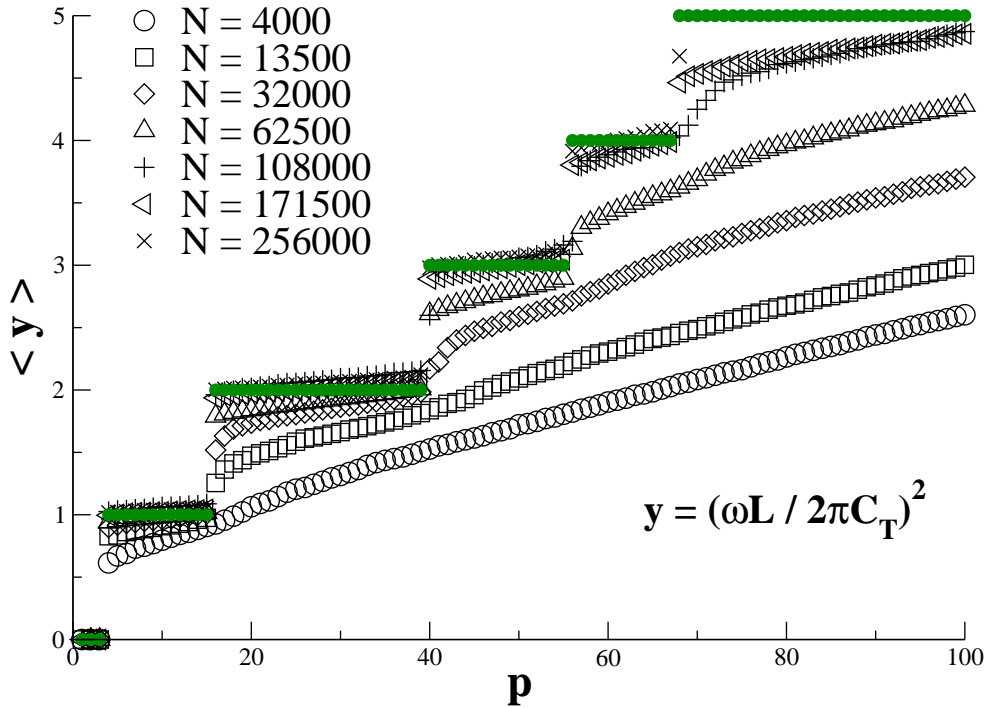}}}
\vspace*{0.0cm} \caption[]{
Eigenfrequencies $\omega$ of the first 100 modes as a function of 
mode number $p$. The frequencies are rescaled using the system size 
and the sound velocity (obtained from the Lam\'e coefficient in 
Fig.~\ref{figlame}). The horizontal lines correspond to the 
results expected from macroscopic elasticity. 
Frequencies of small systems are systematically too low.
Besides the obvious Goldstone modes ($p=1,2,3$), all frequencies
are finite.
\label{figomegap}}
\end{figure}

We turn now to the determination of low frequency vibration modes.
These were determined from a direct diagonalization of the
dynamical matrix, using a modified version of the PARPACK package
\cite{parpack}. For a periodic cubic system described by classical
elasticity, the structure of the low frequency end of the spectrum
is well known. Each mode is characterised by a wave vector $\kvec=
\frac{2\pi}{L} (l,m,n)$.  Transverse  (resp. longitudinal) have
frequencies $\omega = c_T k$  (resp.  $\omega=c_L k$), where the
sound velocity is given by $c_T = \sqrt{\mu/\rho} \approx 4.2$ (resp.
$c_L=\sqrt{(\lambda+2\mu)/\rho} \approx 8.9$).  As a result, the modes should
have well defined degeneracies. For example, the lowest lying mode $(\pm 1,0,0)$
should have 12 fold degeneracy, corresponding to the two
transverse polarisations for the 6 wave vectors of length $2\pi/L$.
The second frequency has degeneracy 24, and so on.  In our
previous analysis of 2D systems \cite{tanguy02}, we
found that this degeneracy of the low frequency modes was lifted
for small systems sizes. A scaling analysis of the modes allowed to
establish in a different manner the length scale $\xi$, above
which the medium can be considered as elastically homogeneous.

Our results for the low frequency modes of three dimensional
systems are shown in Fig.~\ref{figomegap}. 
Plotting the rescaled and averaged frequencies 
$\langle y \rangle = \langle(\omega L/2\pi c_T)^2 \rangle$ 
as a function of mode number $p$,
\footnote{The rescaling eliminates trivial dependencies on system size 
and sound velocities which are anyway expected from continuum theory.
The rescaled frequencies are averaged over the ensemble of available
configurations. Note that differences between the eigenfrequencies of 
different configurations of the ensemble are weak, however, even for 
small system sizes.}
it is clear that the
degeneracies and the associated  step-like behavior begin to show
up only for large systems, containing at least 32000 particles
(lateral size 32). 
For the largest systems however (lateral size
64) the discreteness of the low frequency spectrum is well
apparent, typically up to the 4th eigenfrequencies. In view of the
large value of $c_L$ compared to $c_T$ ($c_L \simeq 2.1 c_T$ in our
system) we have concentrated on the analysis of transverse modes.
Longitudinal modes enter only at higher frequencies, and are mixed
with shorter wave length transverse modes, making their
contribution more difficult to identify. If we take as a criterion
the existence of a gap separating the first 12 eigenfrequencies
from the rest of the spectrum, it appears that the minimum size
for applying continuum elasticity is comprised between $L=16$ and
$L=32$ particle sizes.


\begin{figure}[t]
\centerline{\resizebox{13cm}{!}{\includegraphics*{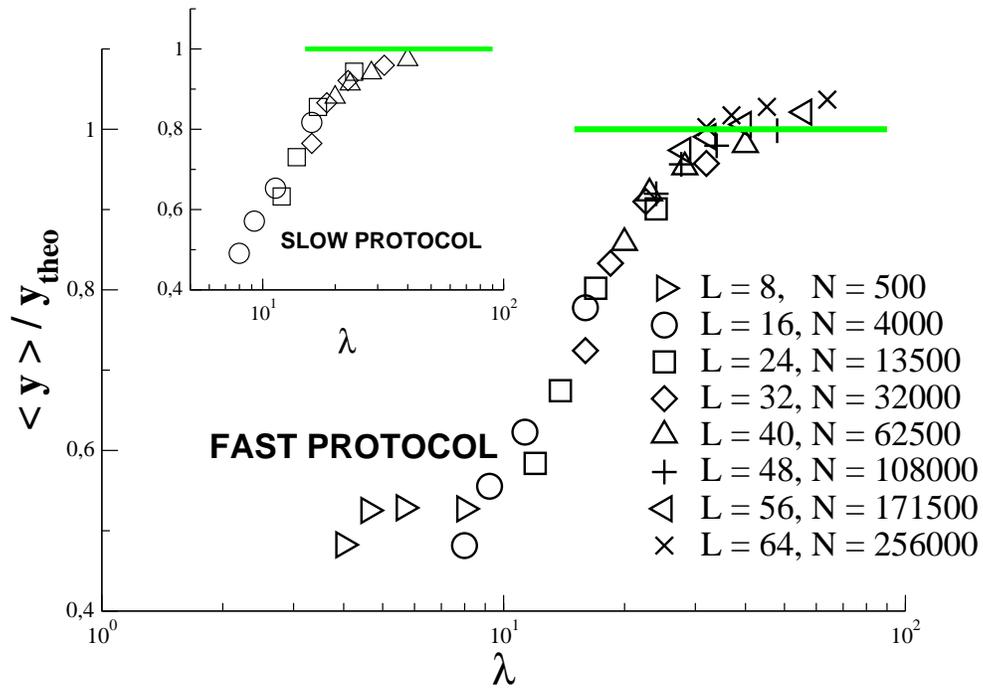}}}
\vspace*{0.0cm} \caption[]{
Eigenfrequencies corresponding to the first four levels predicted by
continuum elasticity, plotted as a function of the corresponding wavelength
$\lambda = 2\pi / \parallel\underline{k}\parallel$.
The eigenfrequencies $\omega$ have been rescaled by their continuum theory
values. This yields a perfect collapse of all data points. 
See main text for details. The crossover to continuum behavior 
(indicated by the horizontal line) takes place, as expected,
at the self-averaging length $\lambda \simeq \xi$. 
The frequencies decrease systematically with smaller $\lambda$. 
Inset: The same plot for the slow quench protocol shows barely
distinguishable behavior (same symbols as in the main plot).
\label{figomegascal}}
\end{figure}

This analysis can be refined using a scaling plot of the mode
frequencies as a function of the ``theoretical'' wave length, or more
precisely the wave length of the elastic wave that would appear in
the spectrum with this mode number according to elastic theory. 
Fig.~\ref{figomegascal} is constructed by averaging, for each size, 
the frequencies that correspond to the
first elastic mode in elastic theory (e.g. the first 12
frequencies are averaged to obtain the lowest frequency point, the
next 24 for the second point, and so on). The resulting frequency,
divided by the value expected from elastic theory, is plotted as a
function of wave length. 
Note that all data points collapse on the same master curve irrespective 
of the box size $L$. The plot shows that when the wave length is
lower than the self-averaging length,
deviations from elastic theory become significant, 
whatever the size of the system. 
This estimate for the size of elastic inhomogeneity is therefore in 
fair agreement with those obtained in two previous sections
from the analysis of the linear response to an external load.

\newpage
\clearpage
\section{Density of vibrational states}
\label{sec:dos}

\begin{figure}[t]
\centerline{\resizebox{13cm}{!}{\includegraphics*{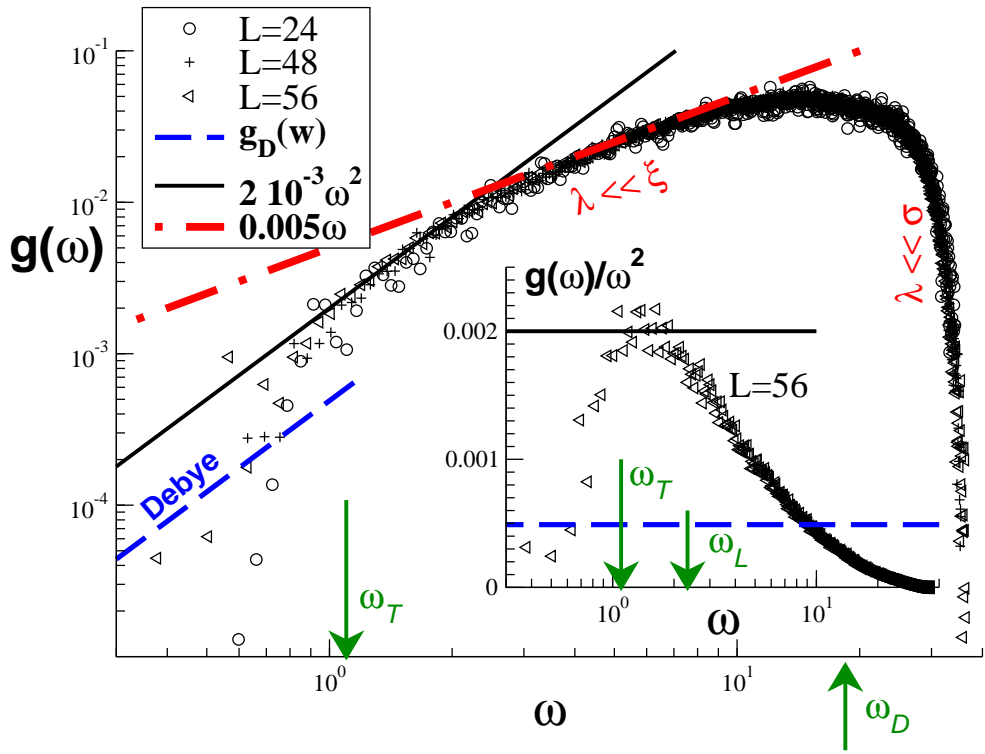}}}
\vspace*{-0.2cm} \caption[]{
Density of states $g(\omega)$ for 3D amorphous systems
of different system sizes $L$.
The lines indicate three power law slopes,
the dashed one being the Debye prediction $g_D(\omega)$
calculated from the known sound velocities.
The dashed-dotted linear relation corresponds to the non-affine
displacement field regime ($\lambda \ll \xi$).
Also given are the characteristic frequencies
$\omega_{L,T} = c_{L,T} 2\pi/\xi$ associated
with the self-averaging length $\xi$ and the
Debye frequency $\omega_D \approx 18.3$.
Larger frequencies correspond, in fact, to vibrations
on very small scales, $\lambda \ll \sigma$.
Inset: $g(\omega)/\omega^2$ {\em vs.} $\omega$
for $L=56$ (same symbols as main figure).
Note that $\omega_T$ and $\omega_L$ describe
correctly the position and width of the boson peak.
\label{figDOS}}
\end{figure}

The (normalized) density of vibrational states (DOS) of a 3D solid
may be defined by $g(\omega) = \frac{1}{3N} \sum_{p=1}^{3N}
\delta(\omega-\omega_p)$ with $\omega_p$ being the harmonic
eigenfrequency corresponding to the mode number $p$. Hence, for
small systems (of order of $10^3$ beads) one can compute the
complete DOS from the eigenfrequencies extracted by exact
diagonalization of the dynamical matrix, just as we have done in
the previous section.
Obviously, for systems containing about $10^5$ particles the
number of modes one may compute is rather limited. From the
100 modes we have presented in Fig.~\ref{figomegap} one estimates
roughly $\omega_p^2 \propto p^{\alpha}$
with $\alpha \approx 1$. Hence, the DOS increases approximately linearly,
$g(\omega) \propto \omega^{2/\alpha-1} \approx \omega$, for small $\omega$.

Following standard procedures \cite{meshkov97}, we have instead
obtained $g(\omega)$ by Fourier transformation of the velocity
auto-correlation $\langle \vvec(t) \cdot \vvec(0) \rangle$. In
contrast to the previous sections, we consider here configurations
at finite, yet very low temperatures $T$. For the data presented
in Fig.~\ref{figDOS} we have used $T=10^{-4}$ which is three
orders of magnitudes below the glass transition of our systems. We
start with quenched configurations at $T=0$ which we subject to a
Maxwell velocity distribution. Following a thermalization phase of
$\delta t=10^3$ the velocity correlation  function is sampled over
$\delta t=100$.
Different temperatures have been checked and we have verified that the
DOS becomes temperature independent at low $T$ (not shown).
As can be seen from the Fig.~\ref{figDOS}, our results become
rapidly system size independent for sufficiently large frequencies
$\omega \gg c_{T,L} 2\pi/L$.  We remember that only $\omega$ values
corresponding to wave vectors $\kvec$ compatible with the box size
are physically acceptable in the continuum limit. The corresponding
finite size effects at low frequencies can be seen on the left hand
side of the figure.

In linear coordinates, $g(\omega)$ is roughly symmetric around its
maximum at $\omega \approx 14.3$ and may be very crudely described
as linear for small $\omega$ in agreement with the estimate from
Fig.~\ref{figomegap} described above and the dashed-dotted line indicated
in the main panel of Fig.~\ref{figDOS}. Note that the maximum
is slightly smaller than the Debye frequency
$\omega_D = \left(18\rho\pi^2/(1/c_L^3+2/c_T^3) \right)^{1/3} \approx 18.3$
for our systems.
(The Debye frequency is in turn smaller than
the frequency $c_T 2\pi/\sigma \approx 26.4$ associated
with a wave length of monomer size.)
The log-log plot presented in the main figure shows various
frequency regimes. For very small frequencies our data
is roughly in agreement with the Debye continuum prediction
$g_D(\omega) = 3 \omega^2 / \omega_D^3$ (dashed line).
The DOS increases than more rapidly with frequency up to
$\omega_T = c_T 2\pi/\xi \approx 1.1$ -- corresponding to
a wave vector given by the self-averaging length -- where
$g(\omega)$ has power law slope of exponent $2$ (bold line).
This can be more clearly seen in the inset featuring the enigmatic
Boson peak. Apparently, the width of this peak is well described by
$\omega_T$ and the frequency $\omega_L \approx 2.3$
for the corresponding longitudinal wave.
Hence, the boson-peak is fixed by the self-averaging length and
marks the crossover between the continuum elastic behavior (dashed
line) and the non-affine displacement field regime, where
$g(\omega) \propto \omega$ (dashed-dotted line), at larger
$\omega$ and smaller wave length $\lambda$. Since continuum theory
overestimates the frequencies
for $\lambda \ll \xi$ 
(see also Fig.~\ref{figomegascal}) this implies an excess
of modes at smaller frequencies. Apparently, these modes
are shifted to the edge of the non-affine regime.
The boson-peak is merely a consequence of the
inapplicability of the continuum theory at $\lambda \ll \xi \approx 23$.
%

\section{Conclusion}
\label{sec:conclusion}

We have investigated the approach of the continuum limit for
elastic properties of 3D amorphous systems and compared our
computational results with our previous work on similar 2D
systems. The results are extremely similar in both cases, and can
be summarized as follows.
The elastic constants estimated using the Born formulae are not accurate
even at zero temperature, therefore revealing the importance of the
non-affine component of the deformation field. This non-affine deformation
field, which affects mostly the shear response (as compared to
compressibility) is correlated over intermediate distances, of the
order of 23 interatomic distances in our case.
This correlation length is significantly smaller than in 2D,
in agreement with the findings of Rossi {\em et al} \cite{rossi},
but implies that rather large samples should be used to discuss
elastic or vibrational properties of 3D systems as well.
By considering the Fourier transformation of solenoidal and longitudinal
part of the non-affine field we have demonstrated (Fig.~\ref{figSTLk})
that the 3D non-affine field is mostly rotational in nature,
in agreement with the visual impression of snapshots.
The response to a delta force perturbation allowed us to
measure the self-averaging of the noisy response within
a configuration. The stress fluctuations decay exponentially
with distance from the source with a characteristic length
given by the correlation length $\xi$ of the non-affine field
which suggests to us the notion ``self-averaging length".

Vibrational modes are obviously strongly affected by the existence
of elastic heterogeneities, and cannot be predicted using elastic
theory if their wave length is too small. From our scaling
analysis (Fig.~\ref{figomegascal}) it appears that the frequencies
are {\em smaller} than expected from continuum theory, therefore
implying an excess of modes in the low frequency region compared
to the Debye prediction. This excess has been analyzed in
Fig.~\ref{figDOS} showing the density of vibrational states and
demonstrating that the ``boson peak" is located at the edge of the
non-affine displacement field.
That both position and width of the peak are given by the
self-averaging length $\xi$ is the central novel result
of this work which has not been previously established for
2D systems.

The focus of this work has been primarily on the {\em linear
elastic} behavior of amorphous solids. Our preliminary study of
larger (uniaxial) deformations that go beyond the elastic limit
indicates that plastic events are rather localized individual
events characterized by a very low participation ratio.
In the recent work \cite{depablo} de Pablo and coworkers pointed
out the possibility of {\em regions of negative shear modulus} in
quenched amorphous systems - such regions being stabilized by the
``normal'' material in which they are embedded. The typical size of
these regions  is much smaller than the size for elastic
inhomogeneities discussed in this work, implying they are more
likely to be linked to elementary  rearrangements taking place at
the onset of plastic deformation, which usually imply small
numbers of particles \cite{maloneylemaitre,falklanger,cavaille,kabla,lacks}, 
or even localization along a shear band \cite{vandembroucq,picard}. 
Such a difference in elastic and plastic deformations was also observed
for 2D systems in Ref.~\cite{maloneylemaitre}.

The general picture that emerges is therefore that of a hierarchy
of length scales. Disorder at the level of 2-3 atomic distances
can be interpreted as implying the existence of regions with
negative moduli, which will give rise to plastic yield. On a
larger scale, this disorder gives rise to strong non-affine
displacement fields in elementary deformations. Finally,
convergence to standard continuum properties is obtained over
length scales larger than the self-averaging length. In
our analysis, carried out for a typical liquid state density and 
at zero-temperature, $\xi$ is found to be large, but finite. 
In analogy with what is found in 2D, we expect it to decrease with increasing
density, and possibly to diverge as the density is lowered and the
system loses mechanical stability, as suggested in Ref.~\cite{nagel}.

\begin{acknowledgments}
During the course of this work, we had valuable discussions with
A.~Lema\^\i tre, C. Maloney, J. de Pablo, B.~Schnell and D.~Vandembroucq.
Computational support by IDRIS/France is acknowledged. F.L. was
supported by the Emergence 2002 Program of the Conseil Regional
Rh\^ones-Alpes (France).
\end{acknowledgments}



\end{document}